\documentclass[11pt]{article}

\usepackage{amsmath}
\usepackage{amsfonts}
\usepackage{amssymb}
\usepackage{graphicx}
\usepackage{supertabular}
\usepackage{setspace}
\usepackage{xcolor}
\usepackage{array}
\usepackage{amsthm}
\usepackage{bm}
\usepackage{latexsym}
\usepackage{mathptmx}
\usepackage{hyperref}
\hypersetup{colorlinks=false}

\RequirePackage{cite}%
\renewcommand{\citeleft}{\bgroup\normalfont[}%
\renewcommand{\citeright}{]\egroup}%

\newcommand{\nin}{\noindent}

\newcommand{\be}{\begin{equation}}
\newcommand{\ee}{\end{equation}}
\newcommand{\ba}{\begin{eqnarray}}
\newcommand{\ea}{\end{eqnarray}}
\newcommand{\bal}{\begin{align}}
\newcommand{\eal}{\end{align}}

\newcommand{\dd}{{\rm d}}

\newcommand{\la}{\lambda}
\newcommand{\bt}{\beta}

\newcommand{\ga}{\gamma}

\newcommand{\Si}{\Sigma}

\newcommand{\Om}{\Omega}

\newcommand{\de}{\delta}

\newcommand{\bw}{\begin{widetext}}
\newcommand{\ew}{\end{widetext}}

\def\abh{black hole }
\def\bh{black holes }

\def\RN{Reissner-Nordstr\"om }
\def\tv{thermodynamic variables}

\begin{document}


\title{{\textbf{\large On `The conformal metric structure of Geometrothermodynamics': Generalizations}}}

\author{Mustapha Azreg-A\"{\i}nou
\\Ba\c{s}kent University, Department of Mathematics, Ba\u{g}l\i ca Campus, Ankara, Turkey}
\date{}

\maketitle

\begin{abstract}
We show that the range of applicability of the change of representation formula derived in J. Math. Phys. \textbf{54}, 033513 (2013) [arXiv:1302.6928] is very narrow and extend it to include all physical applications, particularly, applications to black hole thermodynamics, cosmology and fluid thermodynamics.

\vspace{3mm}


\vspace{-3mm} \nin \line(1,0){430} 
\end{abstract}

We comment on a couple of equations derived in~\cite{conf} and generalize them to apply to a wide range of physical problems pertaining to \abh thermodynamics, cosmology and fluid thermodynamics.

To persuade the reader of the importance of the above-mentioned generalization, we provide two examples from \abh thermodynamics. Consider the thermodynamics of Kerr-Newman \abh governed by the equations (2.3), (2.5), (2.6) to (2.9) of~\cite{d2}:
\begin{align}
\label{1}&M=\{2S+[J^2+(Q^2)^2/4]/(8S)+Q^2/2\}^{1/2}\\
\label{2}&\dd M=T\dd S+\Om\dd J+\phi\dd Q
\end{align}
where $T=\partial M/\partial S$, $\Om=\partial M/\partial J$, $\phi=\partial M/\partial Q$ are given in Eqs. (2.6) to (2.8) of~\cite{d2}. It is easy to check that $M$ is not homogeneous in ($S,J,Q$) for it is not possible to find a real $\bt$ such that $M(\la S,\la J,\la Q)=\la^{\bt}M(S,J,Q)$. The way the right-hand side (r.h.s) of~\eqref{1} has been arranged indicates that $M$ is homogeneous in ($S,J,Q^2$) of degree 1/2:
\begin{equation}\label{3}
  M(\la S,\la J,\la Q^2)=\la^{1/2}M(S,J,Q^2).
\end{equation}
By Euler's Theorem we re-derive Eq. (2.9) of~\cite{d2}:
\begin{align}
\label{4}\frac{1}{2}M=&\frac{\partial M}{\partial S}S+\frac{\partial M}{\partial J}J+\frac{\partial M}{\partial (Q^2)}Q^2\\
\label{5}\quad =&TS+\Om J+\frac{\phi Q}{2}
\end{align}
where we have used $\partial M/\partial (Q^2)=(\partial M/\partial Q)[\partial Q/\partial (Q^2)]=\phi/(2Q)$. It is worth mentioning that the extensive variables ($S,J,Q$) in terms of which the first law~\eqref{2} is written are not the same extensive variables ($S,J,Q^2$) in terms of which the Euler identity~\eqref{4} is expressed. We denote the former variables by $E^a$ and the latter variables by $E^{\,\prime a}$ (in this example: $E^1=E^{\,\prime 1}=S$, $E^2=E^{\,\prime 2}=J$, $E^3=Q$, $E^{\,\prime 3}=Q^2$). The variables $E^{\,\prime a}$ are power-law functions of $E^a$: $E^{\,\prime a}=(E^a)^{p_a}$ (no summation), where $p_a$ depends on $a$. Thus, if $\Phi$ is homogeneous in $E^{\,\prime a}$ of degree $\bt$: $\Phi(\la E^{\,\prime a})=\la^{\bt}\Phi(E^{\,\prime a})$, then the Euler identity reads
\begin{align}
\label{6}\bt\Phi & = E^{\,\prime a}\frac{\partial\Phi}{\partial E^{\,\prime a}}\quad (\Si \text{ over }a,\;a=1,2,\ldots )\\
\label{7}\quad & = \frac{E^a}{p_a}\frac{\partial\Phi}{\partial E^a}\quad (\Si \text{ over }a)
\end{align}
where we have used $\partial E^{\,\prime a}/\partial E^a=p_a(E^a)^{p_a-1}$ (no summation), while the first law is given by
\begin{equation}\label{7b}
    \dd\Phi=I_a\,\dd E^a \quad (\Si \text{ over }a).
\end{equation}
where $I_a=\partial\Phi/\partial E^a$  ($I_a=\de_{ab}I^b$). The authors of~\cite{conf} considered the case where all $p_a\equiv 1$, which is a very restrictive constraint and rarely met in \abh thermodynamics, cosmology or fluid thermodynamics.

Now, if we rewrite~\eqref{1} as
\begin{equation}
    M=\{2(\sqrt{S})^2+[(\sqrt{J})^4+(Q)^4/4]/[8(\sqrt{S})^2]+(Q)^2/2\}^{1/2}
\end{equation}
and regard $M$ as a function of ($\sqrt{S},\sqrt{J},Q$) then $M$ is homogeneous in ($\sqrt{S},\sqrt{J},Q$) of degree 1:
\begin{equation}\label{8}
    M(\la\sqrt{S},\la\sqrt{J},\la Q)=\la M(\sqrt{S},\sqrt{J},Q)
\end{equation}
where $E^{\,\prime 1}=\sqrt{S}$, $E^{\,\prime 2}=\sqrt{J}$, $E^{\,\prime 3}=Q$ with $p_1=p_2=1/2$ and $p_3=1$. By Euler's Theorem we obtain (see~\eqref{6}, ~\eqref{7}):
\begin{align}
\label{9}M=&\frac{\partial M}{\partial \sqrt{S}}\sqrt{S}+\frac{\partial M}{\partial \sqrt{J}}\sqrt{J}+\frac{\partial M}{\partial Q}Q\\
\label{10}\quad =&2TS+2\Om J+\phi Q
\end{align}
where we have used $\partial M/\partial \sqrt{S}=2\sqrt{S}(\partial M/\partial S)$ and $\partial M/\partial \sqrt{J}=2\sqrt{J}(\partial M/\partial J)$. But~\eqref{10} is just~\eqref{5}. This means that (1) one can always choose $\bt =1$ and that (2) the powers $p_a$ depend on $\bt$: $p_a\equiv p_a(\bt)$. If $\bar{p}_a$ denotes the values of the powers for $\bt =1$, then on dividing both sides of~\eqref{7} by $\bt$ one obtains
\begin{equation}\label{11}
    \bar{p}_a=\bt p_a(\bt).
\end{equation}

We can rewrite~\eqref{1} any way we want: If $\ga >0$, we bring it to the form $$M=\{2(S^{\ga})^{1/\ga}+[(J^{\ga})^{2/\ga}+(Q^{2\ga})^{2/\ga}/4]/[8(S^{\ga})^{1/\ga}]+(Q^{2\ga})^{2/\ga}\}^{1/2}$$ where $M$ appears to be homogeneous in ($S^{\ga},J^{\ga},Q^{2\ga}$) of degree $(1/2)/\ga$.

As a general rule: if $f$ is a homogeneous function of ($x,y,\ldots$) of degree $\bt$ then it is a homogeneous function of ($x^{\ga},y^{\ga},\ldots$) of degree $\bt/\ga$. In the special choice $\ga=\bt$, $f$ is homogeneous in ($x^{\bt},y^{\bt},\ldots$) of degree 1.

In another more instructive example consider the thermodynamics of \RN \bh in $d$-dimensions governed by~\cite{RN}
\begin{equation}\label{12}
    M(S,Q)=\frac{S^D}{2}+\frac{Q^2}{4DS^D}\quad \big(D\equiv \frac{d-3}{d-2}\big)
\end{equation}
where
\begin{equation}\label{13}
T=\frac{\partial M}{\partial S}=\frac{DS^{D-1}}{2}-\frac{Q^2}{4S^{D+1}},\;\phi=\frac{\partial M}{\partial Q}=\frac{Q}{2DS^D}.
\end{equation}
It is straightforward to check that $M$ is not homogeneous in ($S,Q$), that is the powers $p_a$ can't all be 1. Assuming $M(\la S^{p_S},\la Q^{p_Q})=\la^{\bt}M(S^{p_S},Q^{p_Q})$ we find
\begin{equation}\label{14}
 p_S(\bt)=D/\bt ,\;p_Q(\bt)=1/\bt ,
\end{equation}
Whatever the value of $\bt$ we choose, it is not possible to have $p_Q=p_S$. If we choose $\bt=1$ this leads to $\bar{p}_S=D$ and $\bar{p}_Q=1$. On applying~\eqref{7} we obtain $M=TS/D+\phi Q$ or $DM=TS+D\phi Q$, which is independent of the choice of $\bt$. It is straightforward to verify the validity of this latter equation upon substituting into its r.h.s the expressions of $T$ and $\phi$ given in~\eqref{13}.

Our starting point is the set of Eqs. (31) to (38) of~\cite{conf} which we intend to generalize. Those equations were derived constraining $\Phi$ to obey the very special Euler identity $\bt \Phi =I_a E^a$ (Eq. (34) of~\cite{conf}). Eqs. (31) to (33) of~\cite{conf} remain valid in the general case~\eqref{7}. Eqs. (35) and (36) of~\cite{conf} have extra misprinted or missing factors. Considering the general case~\eqref{7} with $p_a\neq 1$, the equations generalizing Eqs. (35) and (36) of~\cite{conf} are derived on substituting~\eqref{7}, \eqref{7b}, and Eqs. (32) and (33) of~\cite{conf} into Eq. (31) of~\cite{conf}. They read respectively
\begin{align}
	\label{c1}
 	g^{E^{(i)}} &=  \frac{1}{\beta}  \bigg[ \frac{\xi^{(i)}_{(i)} E^{(i)}}{p_{(i)}}  +  \sum_{j\neq i}\bigg(\frac{\xi^{(i)}_{(i)}}{p_{(i)}} - \xi^j_{j}\beta \bigg) \frac{I_jE^j}{I_{(i)}}\bigg] \nonumber\\
 	&\times \Big[ -\Lambda_{(i)} \chi^{(i)}_{(i)} \frac{1}{I_{(i)}} \,\dd E^{(i)} \otimes \dd I_{(i)} - \Lambda_{(i)} \chi^{(i)}_{(i)} \sum_{j\neq i} \frac{I_j}{I_{(i)}^2} \,\dd E^j \otimes \dd I_{(i)}   \nonumber \\
&-  \sum_{j\neq i} \Lambda_j \chi^j_{j} \frac{1}{I_{(i)}} \,\dd E^j \otimes \dd I_j + \sum_{j\neq i} \Lambda_j \chi^j_{j} \frac{I_j}{I_{(i)}^2} \,\dd E^j \otimes \dd I_{(i)} \Big],
\end{align}
\begin{align}
\label{c2}
 	g^{E^{(i)}} &= \frac{1}{\beta}  \bigg[ \frac{\xi^{(i)}_{(i)} E^{(i)}}{p_{(i)}}  +  \sum_{j\neq i}\bigg(\frac{\xi^{(i)}_{(i)}}{p_{(i)}} - \xi^j_{j}\beta \bigg) \frac{I_jE^j}{I_{(i)}}\bigg]\nonumber \\
& \times \Big[- \sum_{k} \frac{\Lambda_k \chi^k_{c}}{I_{(i)}}  \,\dd I_k \otimes \dd E^c  + \sum_{j\neq i} \left(\Lambda_j \chi^j_{j} -  \Lambda_{(i)} \chi^{(i)}_{(i)}\right) \frac{I_j}{{I_{(i)}}^2}\, \dd E^j \otimes \dd I_{(i)} \Big].
\end{align}
where we have assumed, as in~\cite{conf}, $\chi^k_{c}=\de^k_{c}$ or $\chi^k_{c}=\eta^k_{c}={\rm diag}[-1,1,\ldots ,1]$. Notice the absence of the leftmost factor `$-1/I_{(i)}$' in both expressions and the presence of the same factor in front of the first $\Si$ sign in~\eqref{c2}. The constraints (37) of~\cite{conf} remain unchanged
\begin{equation}\label{c3}
\Lambda_{(i)} = \Lambda_j \,\chi^j_{j}/\chi^{(i)}_{(i)}\,, \; \forall j\neq i\quad (\text{no }\Si \text{ over }j).
\end{equation}
We see that the changes appear in the first factor of each expression only. The final expression of the induced metric, generalizing Eq. (38) of~\cite{conf}, reads
\begin{equation}\label{c4}
  g^{E^{(i)}} = -\frac{1}{\beta I_{(i)}}  \bigg[ \frac{\xi^{(i)}_{(i)} E^{(i)}}{p_{(i)}}  +  \sum_{j\neq i}\bigg(\frac{\xi^{(i)}_{(i)}}{p_{(i)}} - \xi^j_{j}\beta \bigg) \frac{I_jE^j}{I_{(i)}}\bigg][\xi^a_{b} I_aE^b]^{-1} g^{\Phi}.
\end{equation}

As we have seen earlier, we can always choose $\bt =1$, and this choice is not an extra constraint obeyed by some physical systems only as one may infer that to be the case from~\cite{conf}. So, assume that $\Phi$ is homogeneous in some set of \tv, find the set of variables with respect to which $\Phi$ is homogeneous of degree 1 ($\bt =1$), as was done earlier. If the metric $g^{\Phi}$ is chosen such that the constraints~\eqref{c3} are satisfied, and if $\xi^{(i)}_{(i)} = \xi^j_{j}=1$, then the induced metric reduces to
\begin{align}
\label{c5} g^{E^{(i)}} = & -\bigg[ \frac{E^{(i)}}{\bar{p}_{(i)}I_{(i)}}  +  \bigg(\frac{1}{\bar{p}_{(i)}} - 1 \bigg) \frac{\sum_{j\neq i}I_jE^j}{I_{(i)}^2}\bigg](I_aE^a)^{-1} g^{\Phi}\\
\label{c6} \quad = & -\frac{\Phi-\sum_{j\neq i}I_jE^j+\sum_{j\neq i}(\bar{p}_{(i)}^{\ -1}-\bar{p}_j^{\ -1})I_jE^j}{I_{(i)}^2(I_aE^a)}\,g^{\Phi}
\end{align}
generalizing Eq. (53) of~\cite{conf}. Here we have used~\eqref{7} with $\bt =1$: $\Phi=I_{(i)}E^{(i)}/\bar{p}_{(i)}+\sum_{j\neq i}I_jE^j/\bar{p}_j$.

In \abh thermodynamics, if the mass depends only on two extensive variable: $M(S,Z)$ where $Z=Q$ or $Z=J$. In this case, setting $Z=E^j$, $Y\equiv \partial M/\partial Z=I_j$, $E^{(i)}=S$, $I_{(i)}=T$ and $M=TS/\bar{p}_S+YZ/\bar{p}_Z$ in~\eqref{c6} we obtain
\begin{equation}\label{c7}
g^S=-\frac{1}{T^2}\Big[\frac{1}{\bar{p}_S}-\frac{YZ}{TS+YZ}\Big]g^M
\end{equation}
 provided $g^M$ is chosen such that the constraints~\eqref{c3} are satisfied. In the case of \RN \bh in $d$-dimensions, $Z=Q$, $Y=\phi$ and $\bar{p}_S=D$.

Our next point is to generalize Eqs. (51) and (52) of~\cite{conf}. These two equations have been derived from Eq. (50) of~\cite{conf} on substituting $\Phi$ by the special form $I_aE^a/\bt$. Their generalizations are straightforwardly derived on substituting in Eqs. (51) and (52) of~\cite{conf} $I_aE^a/\bt$ by $I_aE^a/(\bt p_a)=I_aE^a/\bar{p}_a$ [Eq.~\eqref{7}].


\end{document}